\documentclass[a4paper]{jpconf}
\usepackage{graphicx} 
\usepackage{multirow}%
\usepackage{amsmath,amssymb,amsfonts}%
\usepackage{amsthm}%
\usepackage{mathrsfs}%
\usepackage[title]{appendix}%
\usepackage{xcolor}%
\usepackage{textcomp}%
\usepackage{booktabs}%
\usepackage{algorithm}%
\usepackage{algorithmicx}%
\usepackage{algpseudocode}%
\usepackage{changepage}
\usepackage{amssymb,amsmath}
\usepackage{mathrsfs}
\usepackage{txfonts}
\usepackage{subcaption}
\usepackage{comment}

\begin{document}
\title{Development of a laser stabilized on an ultra-stable silicon cryogenic Fabry-Perot cavity for dark matter detection}

\author{Yara Hariri$^{1}$, Jacques Millo$^{1}$, Clément Lacroûte$^{2}$, Joannès Barbarat$^{2}$, Yann Kersalé$^{1}$ and Jonathan Gillot$^{1}$}

\address{
$^{1}$ENSMM, CNRS, FEMTO-ST, 26 rue de l’Epitaphe, 25000 Besançon, France\\
$^{2}$Université de Franche-Comté, CNRS, FEMTO-ST, 26 rue de l’Epitaphe, 25000 Besançon, France\\
}
\ead{jonathan.gillot@femto-st.fr}
\vspace{10pt}
\begin{indented}
\item[]October 2023
\end{indented}
\begin{abstract}
    Astrophysical observations suggest the existence of an unknown kind of matter in the Universe, in the frame of the $\Lambda$CDM model. The research field of dark matter covers an energy scale going from massive objects to ultra-light scalar fields, which are the focus of the present work. It is supposed that ultra-light scalar fields affect the length of objects, whereas the speed of light stays unchanged. It follows that Fabry-Perot cavities are ideal tools for ultra-light dark matter detection since the fluctuations in the length of a cavity can be detected on the frequency of the laser stabilized to it. At FEMTO-ST, we have set up an ultra-stable silicon cavity suitable for a test of detection of ultra-light dark matter in an energy range close to 10$^{-10}$ eV. Our 14 cm cavity is composed of two mirrors optically bonded to an ultra-rigid spacer, with each element made in single-crystal of silicon, and cooled at 17 K in order to cancel the first order thermal expansion coefficient of the silicon spacer. The projected fractional frequency stability of the laser is $3 \times 10^{-17}$, mainly limited by the thermal noise of the amorphous dielectric reflective coatings. To reach this remarkable stability, several effects have to be reduced below the thermal noise limit. While the contribution of the residual amplitude modulation is now acceptable, we are currently implementing a laser power lock with residual fluctuations lower than 3~nW and a piezoelectric-based servo loop to actively reduce the vibration noise that has to be inferior to $-110 \textrm{~dB(m s}^{-2})^2/\textrm{Hz}$ at 1 Hz. Here, we present both the status of the development of our ultra-stable laser and the mechanical response of the cavity in the presence of ultra-light dark matter.
\end{abstract}

\section{Introduction}

It has been over 30 years since ultra-stable lasers based on Fabry-Perot cavities have been proposed and developed \cite{salomon_laser_1988}. With the increasing refinement of both material choices, spacer geometry, and overall design, lasers with fractional frequency instabilities of 10$^{-16}$ and below have been developed  \cite{matei_15_2017-1}, \cite{Robinson19}.

The advantages of using crystalline materials at cryogenic temperatures for both the spacer and the mirror substrates were pointed out pretty early on \cite{sulc_temperature_2003-4}; the combination of low mechanical loss materials, low-temperature operation, and, in some cases, the cancellation of the linear thermal expansion coefficient, ensure both a very low thermal noise and a low long-term drift. With the advent of closed-cycle cryostats, the continuous operation of such cavities was an additional advantage identified by several national metrology institutes \cite{zhang_ultrastable_2017-3}, \cite{wang_design_2023}.

 The extremely low thermal noise-induced fractional frequency instability flicker floor of such resonators is indeed unrivaled, making them the preferred choice as local oscillators for optical atomic clocks \cite{nicholson_systematic_2015-1}, \cite{Brewer19}. A 21~cm long silicon cavity cooled to 124 K recently achieved the remarkable fractional frequency stability of $\sigma_y(\tau) = 4 \times10^{-17}$ at 1 s \cite{Matei17}. On longer time scales, ultra-stable cryogenic cavities now exhibit very small frequency drifts over time \cite{Zhadnov19}, with a frequency drift of $3 \times10^{-19}$/s in \cite {Robinson19}. Additionally, ultra-stable Fabry-Perot resonators are particularly suitable for fundamental physics tests, such as ultralight dark matter detection, gravitational waves detection \cite{drever_1981}, and research on Lorentz invariance violation \cite{antonini_test_2005}. In particular, for ultralight dark matter detection, thanks to the interaction between the oscillating scalar field and several fundamental constants that induce fluctuations in the length of the cavity, the oscillating scalar field can then be detected by measuring the frequency fluctuations of the laser-stabilized on the cavity.

In this article, we present the ongoing work in the development of an ultra-stable single-crystal silicon cavity. We start by presenting the experimental setup and discussing the limiting effects on the stability of the cavity and the servo loops implemented to reduce them. We then discuss the mechanical response of the cavity in the presence of an ultra-light dark matter field, emphasizing the amplification of this effect due to the high mechanical quality factor of silicon when compared with ULE glass or fused silica.

\section{Ultra-stable cryogenic Fabry-Perot cavity}
The heart of the setup is a single-crystal silicon cavity, made of a horizontal tapered spacer and two optically contacted mirrors with single-crystal silicon substrates and amorphous dielectric coatings. The optical axis is 140 mm long, and we have measured a finesse of 78000. We attribute this rather low finesse to the surface quality of the mirror substrates. 
At 17 K, the predicted thermal noise of such a cavity is estimated to be at 3 $\times$ 10$^{-17}$.

To reach this level, several factors limiting the frequency stability of the laser should be compensated or stabilized. The most notable sources of noise in the optical cavity are vibrations, temperature fluctuations, and optical power fluctuations. Other instabilities originate from the residual amplitude modulation (RAM), the Doppler effect, and detection noise.

\subsection{Cryogenerator}
\begin{figure}[h!]
    \centering
    \includegraphics[width=0.9\textwidth]{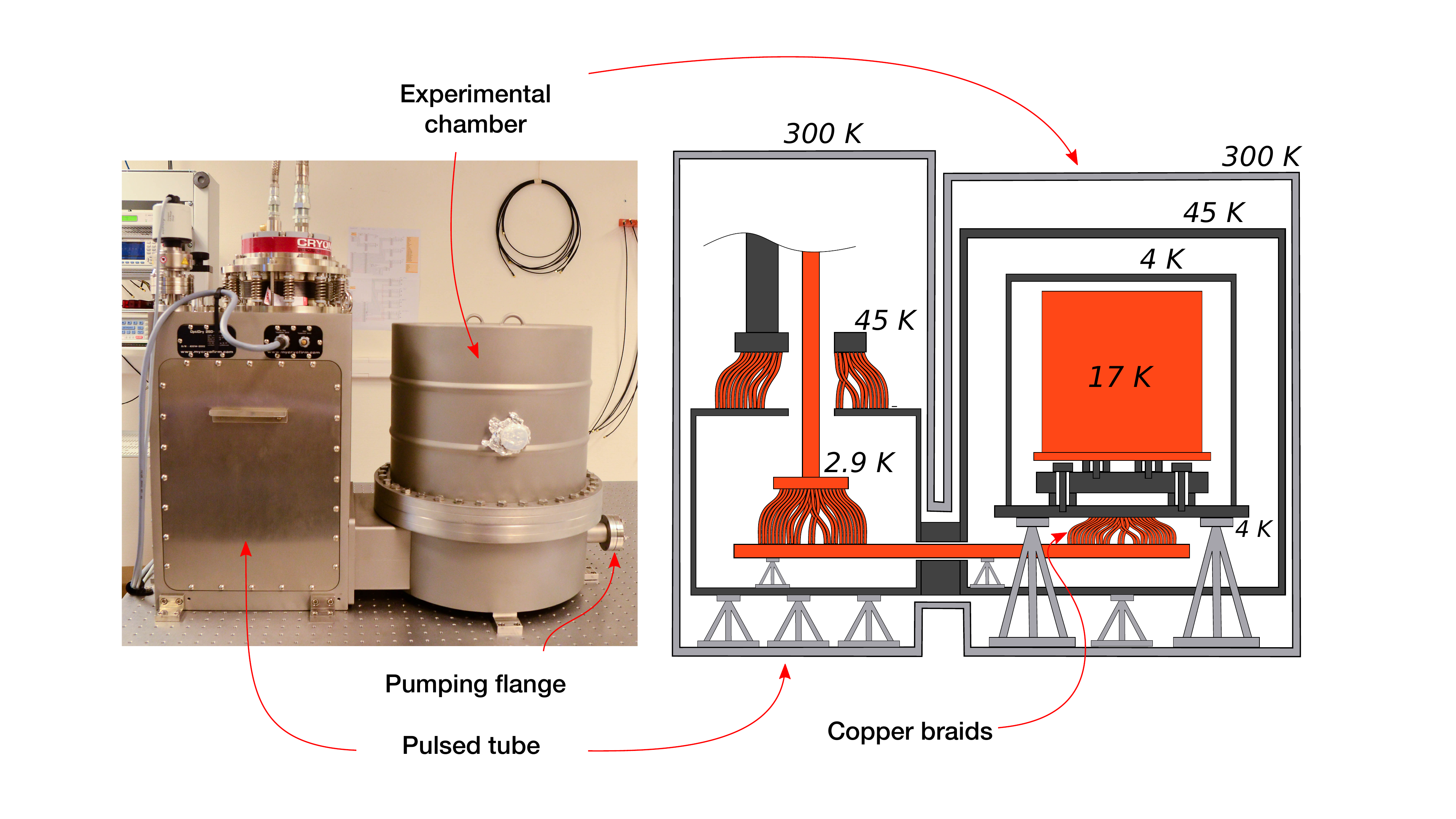}
    \caption{Setup of the two-part cooling system with a pulsed tube cryogenerator.  }
    \label{cryo}
\end{figure}

With a custom-made closed-cycle cryostat, the cavity is cooled down to $\sim$17 K \cite{falzon_2022} to null out the first-order coefficient of thermal expansion.
The cryostat that houses the Fabry-Perot cavity was designed by taking into account three constraints: a good isolation from the cryo-cooler vibrations, a range of operating temperatures from 4 K to 25 K, and a good temperature stability at the silicon temperature sensitivity turnover point at 17 K. To do so, the experimental chamber which holds the cavity inside of it, is separated from the cooling chambers, where the cryo-coolants arrive as seen in Figure~\ref{cryo}. 
The whole setup enclosing the cavity is placed inside a vacuum chamber, to limit the fluctuations of the index of refraction which in return induce fluctuations of the cavity length.

Cooling is ensured by a Cryomech PT410 pulse tube and two-stage cooling, with a first stage at $\sim$45 K, cooling a first thermal shield, and a final stage is a little below 4 K. Thermal transfer between the two chambers is performed by an OFHC copper guide. It is connected to the 4 K cooling stage using loose copper braids that provide both good thermal conductance and vibration isolation.

In the experimental chamber, the baseplate is supported by three rigid tripods. In this way, the experimental baseplate is “mechanically grounded” to the optical breadboard, which itself rests on active vibration-isolation posts. 
The mechanical contact between the cryocooler head which is the main source of vibrations and the experimental chamber  is done only by the loose copper braids.

The experiment baseplate supports a 4 K thermal shield, itself housed inside a 45 K thermal shield. 
A final thermal shield is regulated with the cavity inside at its thermal expansion turnover point. To do so, four heating elements (100~$\Omega$, 10 W resistors) associated with four temperature monitors NTC10K, are attached around a cylindrical support made of gold-coated OFHC copper, and are in contact with this final shield.

The temperature fluctuations are of 1.4 $\times$ 10$^{-4}$ K at 1 s integration time, which corresponds to induced frequency fluctuation of $\sigma_y (\tau)=3 \times 10^{-19}$ at 1 s integration time \cite{falzon_2022}. Therefore the contribution of the temperature fluctuations on the frequency stability is well below the thermal noise limit of 3 $\times$ 10$^{-17}$, which is essentially due to the thermal noise of the reflective coatings \cite{didier_developpement_2016}.

\subsection{Optical setup}

\begin{figure}[h!]
    \centering
    \includegraphics[width=1\textwidth]{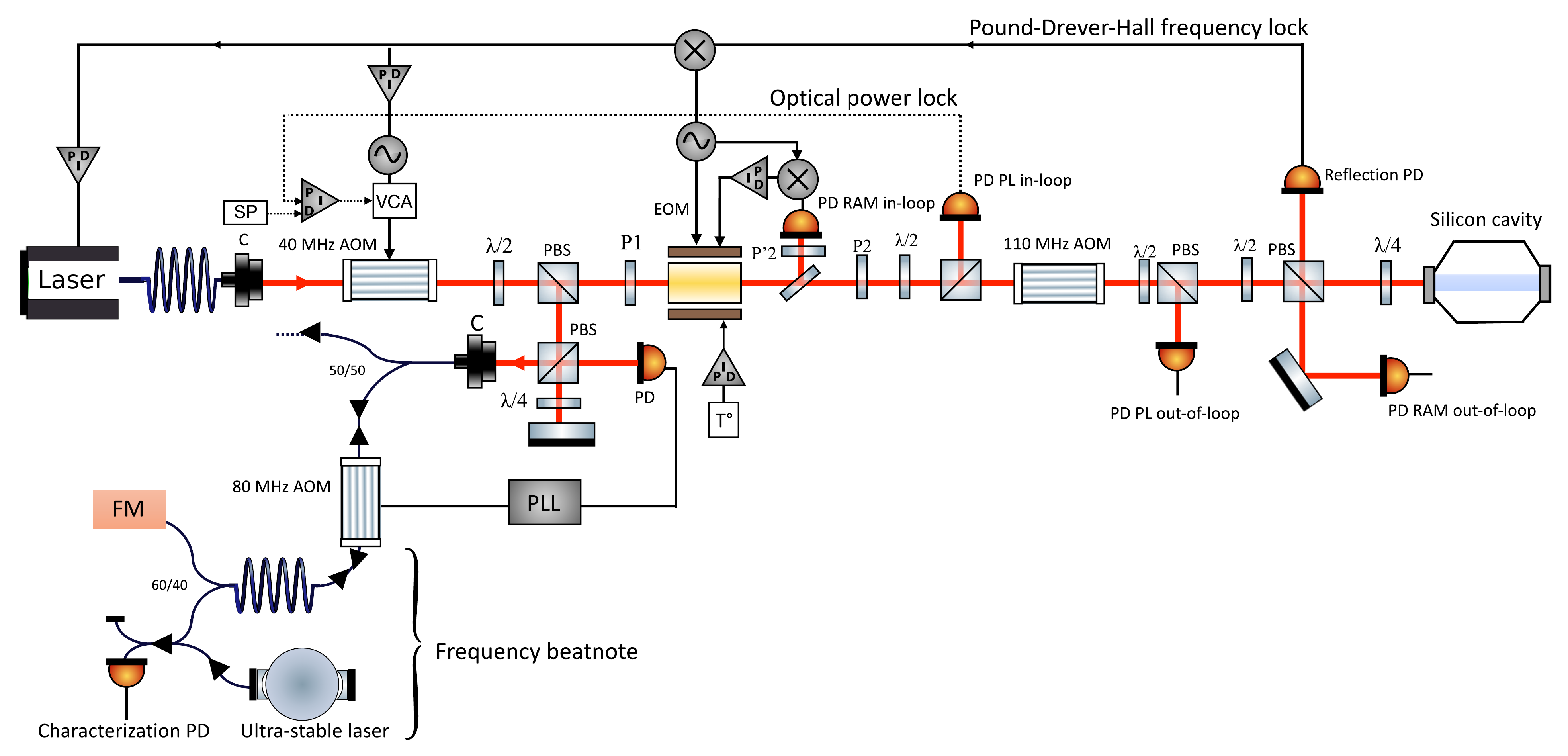}
    \caption{Set-up for stabilizing the frequency of the laser to the Fabry-Perot cavity. EOM: electro-optical modulator, AOM: acousto-optic modulator, $\lambda$/2: half-wave plate,  $\lambda$/4: quarter-wave plate, PBS: polarizing beam splitter, FM: Faraday mirror, C: collimator, P: polarizer, PLL: phase-locked loop, PD: photodiode, PL: power lock, SP: setpoint, VCA: voltage-controlled attenuator. }
    \label{scheme}
\end{figure}

The whole experimental scheme is shown in Fig.\ref{scheme}. The laser source emits light at 1542~nm in free space towards a 40 MHz acousto-optical modulator (AOM), serving the dual purpose of rapid frequency corrections and compensating for power fluctuations in the laser. After that, the laser beam undergoes phase modulation through an electro-optical modulator (EOM) used for the Pound-Drever-Hall (PDH) \cite{black_introduction_2001} frequency stabilization method, resulting in the generation of two sidebands. The modulated light is directed to a 110 MHz AOM, which will be later used to compensate for the Doppler effect before being directed toward the cavity. 
We lock the laser to the TEM$_{00}$ resonance mode of the cavity using the PDH method. The cavity reflection is detected by a fast photodiode as seen in the scheme of Fig.~\ref{scheme}.
The demodulated signal is then the PDH error signal, which is fed to a numerical control loop filter that acts on a piezo-transducer for the long-term corrections, and on the 40 MHz AOM for the short-term corrections.

A drawback of this scheme is the residual amplitude modulation (RAM), which significantly affects the frequency stability of the laser by introducing an uncontrolled offset on the error signal. This effect arises primarily from the polarization mismatch between the polarization plane of the incident light and the extraordinary axis of the EOM crystal \cite{jin_suppression_2021-4} and from stray etalon effects \cite{dangpeng_weak_2010-8}. The suppression of the RAM is done by combining the stabilization of the EOM temperature near cancellation of the RAM and an active servo loop acting on the EOM DC input as shown in  Figure~\ref{scheme} \cite{gillot_digital_2022}. 

The temperature of the EOM is controlled within $\pm$20 mK using a Peltier device. The RAM signal acquired from the in-loop photodiode (PD1) at 22.9 MHz is digitally processed. To derive the digital error signal, precise phase adjustments are made during demodulation. After being processed by a PI controller, the output correction signal is then applied to the DC port of the EOM to control its voltage. The bandwidth of the RAM servo loop is close to 8 kHz.
With this technique, the RAM contribution to the fractional frequency instability is at a level of 5 $\times$ 10$^{-19}$, significantly below the thermal noise limit of the cavity at  3 $\times$ 10$^{-17}$.

Furthermore, the fluctuations of the laser power injected in the cavity result in temperature fluctuations of the mirrors, hence inducing length fluctuations of the cavity. In addition, fluctuations of the laser power induce temperature effects in the EOM crystal leading to an increase in the RAM effect. 

\begin{figure}[h!]
    \centering
    \includegraphics[width=\textwidth]{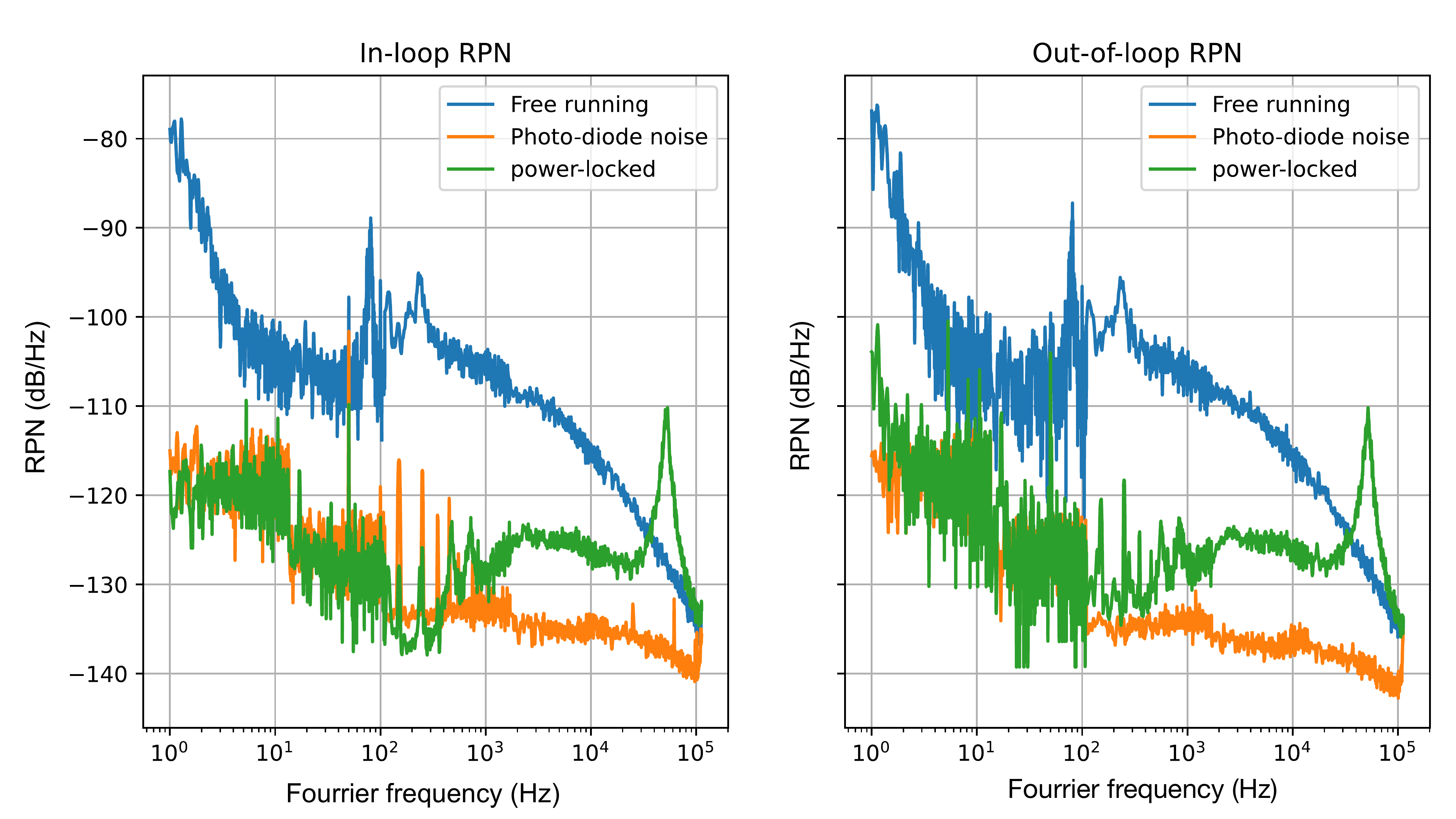}
    \caption{In-loop and out-of-loop RPN respectively.}
    \label{power_lock}
\end{figure}

In order to control these laser power fluctuations, the signal acquired from the photodiode PD PL in-loop shown in Figure~\ref{scheme} is compared to an adjustable voltage (SP) issued from a voltage reference, obtaining thus an error signal. The error signal is fed to a proportional-integrator (PI) controller, that delivers a correction signal to a voltage-controlled attenuator (VCA), which regulates the radio-frequency (RF) power sent to the 40 MHz AOM, thereby stabilizing the optical power. To assess the performance of the power lock in the frequency domain, we performed an analysis of the noise power spectral density (PSD) using a Fast Fourier transform analyzer. The results of both in-loop and out-of-loop measurements are presented in Figure~\ref{power_lock}. The system maintains a relative power noise (RPN) below -100 dB/Hz until 10$^5$ Hz, for both in-loop and out-of-loop. As seen in Figure~\ref{power_lock}, the locked regime noise level for both in-loop and out-of-loop are limited by the photodiodes noise. 

\begin{figure}[h!]
    \centering
     \includegraphics[width=12cm]{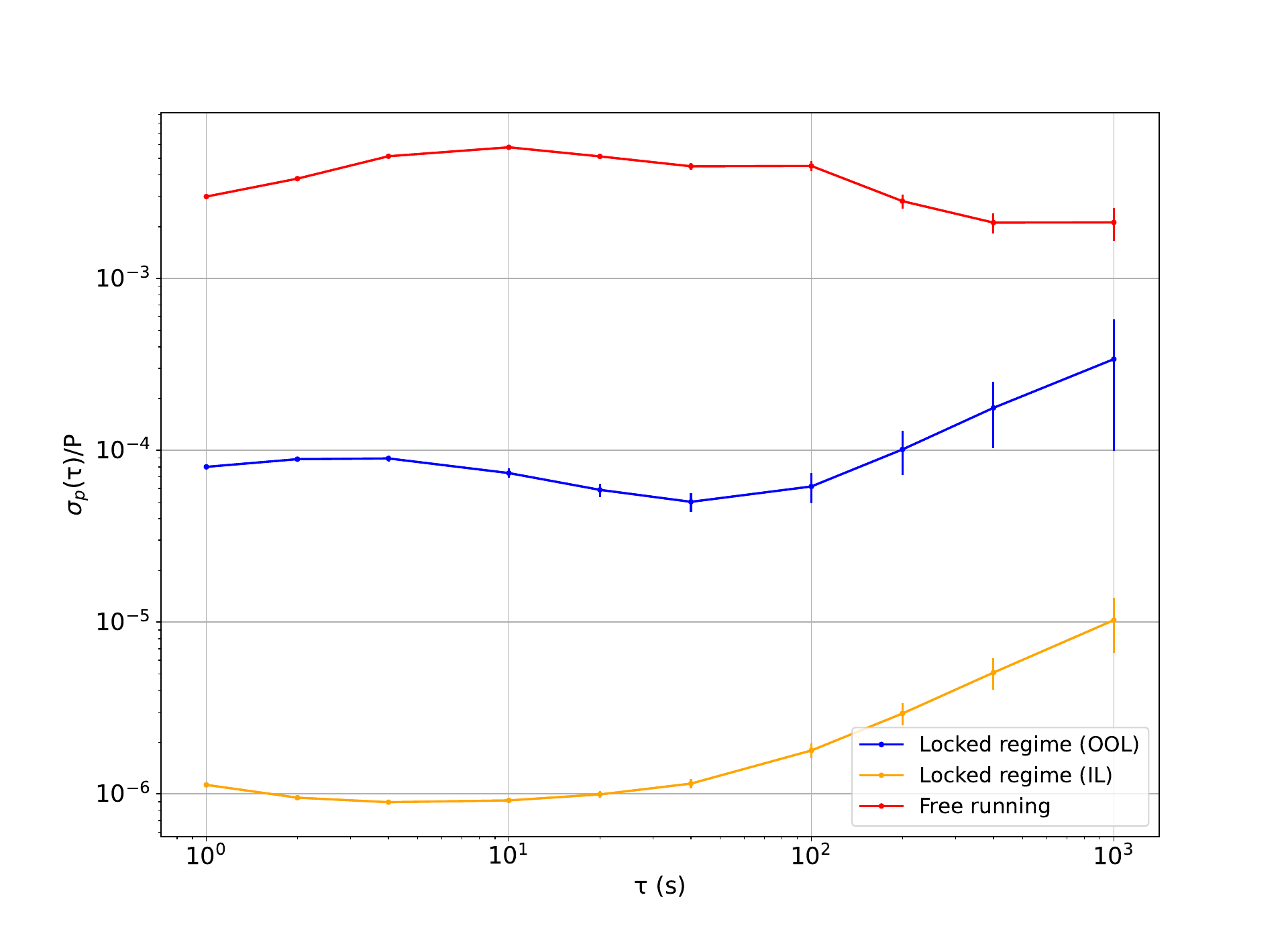}
    \caption{Relative laser-power instability. Red: free running regime, orange: in-loop locked regime, blue: out-of-loop locked regime.}
    \label{power_lock_ADEV}
\end{figure}

To estimate the slow variations of the optical power lock in the time domain, we perform an Allan deviation measurement. The results are presented in Figure~\ref{power_lock_ADEV}. The measured fractional laser-power instability from 1 s to $\sim$100 s on the in-loop PD is around 10$^{-6}$ whereas for the out-of-loop PD is  $\sim$ 10$^{-4}$. 
To determine if the level of the power lock is enough to limit the frequency instability caused by fluctuations of the optical power inside the cavity, it is necessary to measure the sensitivity of the laser frequency on the optical power. Preliminary measurement of this coefficient was done by modulating the optical power using the 40 MHz AOM in the setup shown in Figure~\ref{scheme} using a VCA and monitoring the fractional frequency of the beatnote. The sensitivity coefficient is approximately 2 Hz/$\mu$W. The optical power contribution should be below the thermal noise limit of the cavity $\sigma_y^{th}$=3$\times$10$^{-17}$, which implies the necessity of stabilizing the optical power at the level of a few nanowatts. Therefore, with the power lock, the contribution of the optical power fluctuations to the fractional frequency stability of the cavity is $\sigma_y^{p}$ $\approx$10$^{-18}$ well below the thermal noise of the cavity.

However, the modulation of the optical power was done with an AOM positioned upstream of the EOM, and the power fluctuations within the EOM increased the RAM  \cite{gillot_digital_2022}. Further measurements on this sensitivity coefficient will be conducted, excluding the influence of the RAM, by stabilizing the optical power and applying the modulation on another 110 MHz AOM placed downstream of the EOM.

\subsection{Vibration sensitivity}

One of the major constraints on the ultra-stable laser frequency stability is the environmental vibration noise from acoustic and seismic origins. These vibrations lead to elastic deformations of the cavity that modulate the spacer length. Vibrations-induced frequency noise on the laser locked to the resonance frequency of the cavity is modeled as:

\begin{equation}
    S_{\nu}(f) = {\nu_0}^2 \sum_{i \in \{x,y,z\}} \left| k_i(f) \right|^2 \, S_{a_i}(f),
    \label{vibrations_Eq}
\end{equation}

where $S_{a_i}(f)$ represents the acceleration noise PSD in each direction $x,y,z$ with $f$ the Fourier frequency and $\nu_0$ the frequency of the laser. Therefore, the total frequency noise due to accelerations of the cavity is the sum of the contributions along each direction weighted by the modulus squared of the complex coefficients of sensitivity to vibration $k_i(f)$ of the cavity.

Experimental evolution of these coefficients is often obtained by measuring a transfer function by applying vibrations to the optical table supporting the cavity within its vacuum system and recording the induced frequency noise or fluctuations. So, the coefficients measured include the response of the holding support of the cavity and the vacuum chamber that shows mechanical resonances. In a first approximation, we can consider the coefficients of sensitivity to vibrations constant and independent of $f$ below the first resonances assumed to be higher than 10~Hz.

\begin{figure}[h!]
    \centering
    \includegraphics[width=\textwidth]{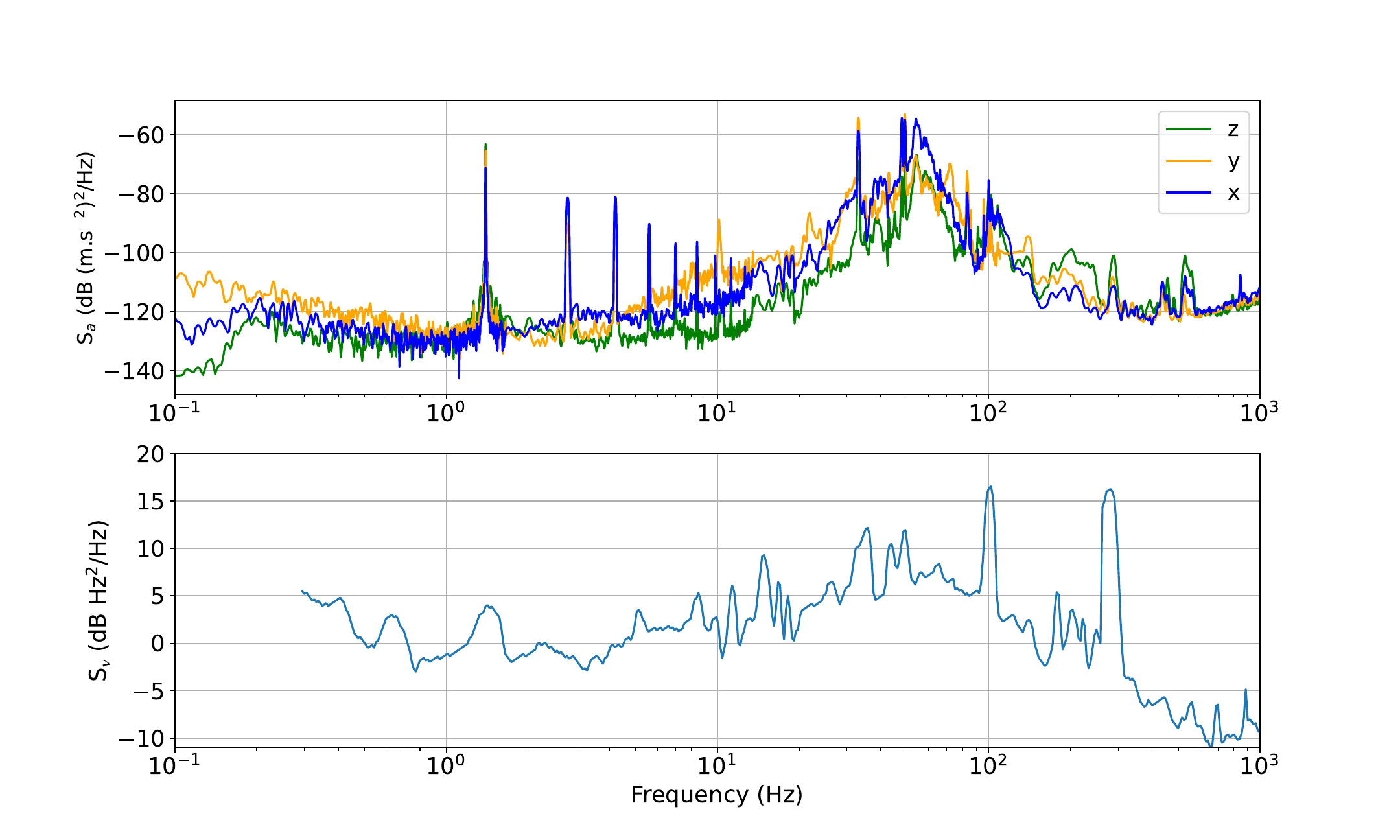}
    \caption{Up: vibration PSD measurements along the 3 directions using a seismometer placed on the optical table next to the cavity when the cryostat is turned on. $z$-axis is the vertical direction, $y$-axis is the horizontal axis along the optical axis of the cavity and $x$-axis is the other horizontal axis, orthogonal to the cavity axis. Down: frequency noise of the cryogenic silicon cavity stabilized laser beating with an ultra-stable laser. }

    \label{vibrations}
\end{figure}

Vibrations rising from the ground are attenuated by an active vibration isolation platform that supports an optical table with the optical setup and the cryostat. Vibrations from the cryostat are minimized by (i) the limited mechanical coupling between the pulse tube, responsible for a large source of vibrations, and the vacuum chamber and (ii) attaching the pulse tube head, which stands outside the cryostat, to a robust and rigid frame surrounding the entire setup. A seismometer with a bandwidth from 17~mHz to 100~Hz placed on the optical table measures the acceleration noise PSD shown in Figure~\ref{vibrations} when the cryocooler is turned on. The noise floor is similar for each direction around 1~Hz with a level of about -130~dB(m s$^{-2})^{2}$/Hz. The noise strongly increases in the 10~Hz to 100~Hz decade due to the high level of acoustic perturbations. 
At very low frequencies, about 0.1~Hz, a difference of 40~dB is observed between the vertical $z$-axis and the horizontal $y$-axis parallel to the optical axis of the cavity. The difference is attributed to the mass distribution on the optical table that has been empirically optimized to get the lowest global vibration noise. 
At $\sim$1.4~Hz and multiple frequencies, we also observe narrow lines corresponding to the harmonics of the vibration induced by the pulse tube. The acceleration PSD value in each direction at the pulsed tube frequency is between -63 and -71 dB(m~s$^{-2})^{2}$/Hz. 

To estimate the performance of the silicon cavity-stabilized-laser, a beatnote between the laser locked to the silicon cavity and another laser locked to a spherical ultra-low-expansion glass (ULE) cavity is detected with a fast photodiode. This laser locked to the ULE cavity achieves a frequency stability of $\sim 2\times 10^{-15}$ at 1~s integration time \cite{didier_developpement_2016}. The frequency noise of the stabilized laser shows a peak at $\sim 1.4$~Hz with a strength of $\sim 4 \textrm{~dB Hz}^2/\textrm{Hz}$ (see Figure \ref{vibrations}).

Using the vibrations noise measured and presented in the previous paragraph we can estimate an upper bound of each $\left|k_i\right|$ coefficient value assuming that they have the same magnitude and contribute with the same phase. The estimated value is $\sim 3.4 \times 10^{-11}/(\textrm{m s}^{-2}$), higher by a factor $\sim 7$ than the sensitivity determined with finite elements modeling but this sensitivity coefficient has not been experimentally optimized. To achieve a frequency instability limited by the thermal noise of this 17~K cavity, a more accurate measurement and an optimization of these coefficients are required. The level of vibrations from the cryostat at $\sim1.4$~Hz and its harmonics, has also to be strongly reduced with feedforward techniques or with an active compensation using high sensitivity seismometer.


\subsection{Preliminary fractional frequency stability measurements}

\begin{figure}[h!]
    \centering
    \includegraphics[width=0.9\textwidth]{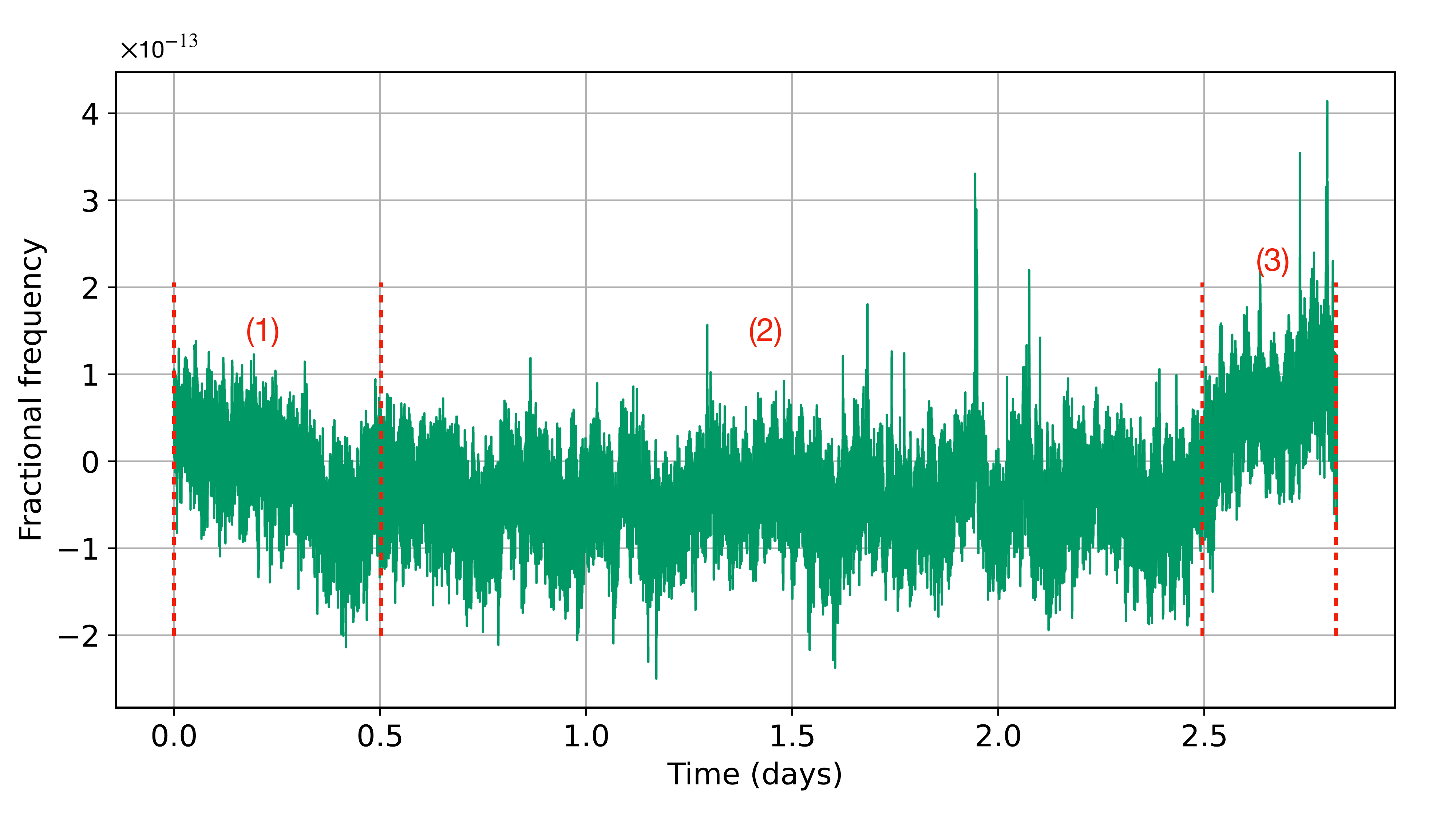}
    \caption{Fractional beatnote frequency between the laser locked to the silicon cavity and the laser locked to the spherical cavity over 3 days. The dataset is divided into 3 sections which will be useful to estimate the drift of the cavity. }
    \label{frac_freq_sicav}
\end{figure}

To determine the frequency drift of the silicon cavity, we first compensate for the drift of the ULE cavity by steering the phase of the optical signal to the phase of an ultra-stable RF reference (active hydrogen maser). The phase comparison is performed in the RF domain using an optical frequency comb as a divider for the optical signal \cite{gushing2023}. We recorded the beatnote frequency shown in Figure~\ref{frac_freq_sicav} for about 3 days. Large frequency changes of $\sim1.5\times10^{-13}$ to $\sim2\times10^{-13}$ between the de-drifted ULE stabilized laser and the silicon stabilized laser in half a day are observed at the beginning and the end of the dataset (respectively section (1) and (3) on Figure~\ref{frac_freq_sicav}). These large variations are attributed to unstable environmental conditions and an excessive sensitivity of the experimental setup. During the two days corresponding to section (2), no large frequency changes or trend is observed. We attribute this behaviour to a lower instability of the environmental parameters.

\begin{figure}[h!]
    \centering
    \includegraphics[width=\textwidth]{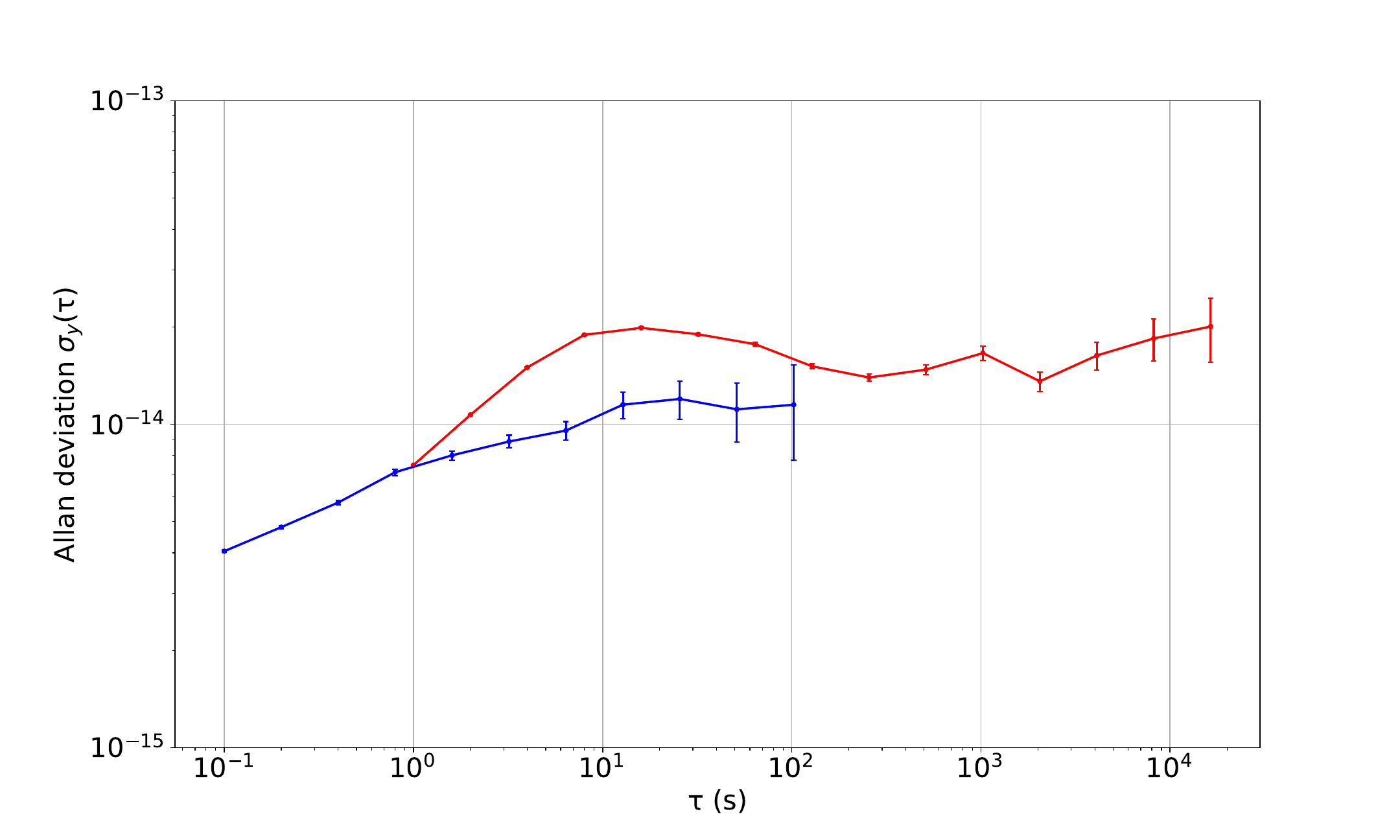}
    \caption{Allan deviation of the beatnote signal between the laser stabilized to the silicon cavity and an ULE reference laser. The red curve represents the fractional frequency instability of the signal shown in Figure~\ref{frac_freq_sicav}. The blue curve is the best instability measured for the same beatnote signal.}
    \label{long_stab_sicav}
\end{figure}

The fractional frequency drift extracted from this set of data is ranging from $\sim 7\times10^{-20}/$s to $4.4\times10^{-18}/$s depending on the considered sections. The fractional frequency stability corresponding to the full data set is reported in red on Figure~\ref{long_stab_sicav} and is between $10^{-14}$ and $2 \times 10^{-14}$ from few seconds integration time to $\sim 10^{4}$~s. No frequency drift is visible on that measurement but an upper bound of about $10^{-18}/$s is compatible with this set of data and with the possible drift rate extracted from the data shown in Figure~\ref{frac_freq_sicav}. This upper bound of $10^{-18}/$s is compatible with the state-of-the-art cryogenic silicon cavities exhibiting fractional frequency drift of about $10^{-19}/$s \cite{ptb_4k, ptb_124k}.

The short term fractional frequency instability of the beatnote signal is $4\times 10^{-15}$ at 0.1~s (blue curve on Figure~\ref{long_stab_sicav}). The blue curve is computed using another measurement in order to estimate the current best performance fractional frequency stability. The progress made thus far has resulted in a frequency stability of the laser locked on the silicon cavity of 4 $\times$ 10$^{-15}$ at 0.1~s integration time through the implementation of the RAM and temperature lock. The power lock at the time of these stability measurements was still not operational. However, there remain various limiting factors that degrade the stability of the laser which include vibrations, Doppler effect, and detection noise as mentioned previously.




\newpage
\section{Dark matter detection}

Dark matter is a central question in fundamental physics and the scientific community has still not succeeded in revealing its origins and composition. This substance leaves indirect traces of its existence, and through gravitational interaction \cite{Zwicky33,Rubin70,Rubin80} its concentration in the Galaxy is estimated at $\rho=0.4$ GeV/cm$^3$ \cite{Mcmillan11}, but it has never been directly detected up to now \cite{Bertone18}. Since the constituents of dark matter can have a mass between $10^{-33}$ GeV and $10^{48}$ GeV \cite{Safronova18}, there is a plethora of models and of experiments to try to determine its nature and properties \cite{Bertone18}.

Among all the scenarios, the study of dark matter has focused a lot on the direct detection of massive constituents between 10 and 1000 GeV, WIMPs (weakly interacting massive particles), without success \cite{Safronova18}. Another possible candidate, the ultra-light dark matter, is now the subject of increased research particularly since the discovery of the Higgs boson at the Large Hadron Collider \cite{Higgs64,Englert64,AtlasC12,CMS12a}. Since this discovery showed the existence of a scalar field to explain the mechanism giving mass to the baryonic particles, the search for dark matter in the form of a scalar field is experiencing renewed activity. Indeed, in this model, dark matter is represented in the form of an oscillating scalar field, and this oscillation is imprinted on the values of the fundamental constants of nature \cite{Arvanitaki15,Stadnik15,Damour10,Hees18}. Thanks to the continuous improvement of performances in time-frequency metrology, attempts to directly detect dark matter with atomic clocks or ultra-stable cavities are increasing \cite{Arvanitaki15,Stadnik15,Derevianko14,Vantilburg15,Hees16,Wcislo17,Roberts17,Wcislo18,Roberts21,Wolf19}.

\subsection{Interaction between the ultrastable cavity and a scalar field}
In this theory, a scalar field $\phi$ of mass $m_\phi$ is linearly coupled to the Lagrangian of the Standard Model, with $m_\phi \ll 1$ eV. An oscillation of the scalar field will produce a variation in the value of several fundamental constants, namely the fine structure constant $\alpha$, the constant $\Lambda_3$ linked to quantum chromodynamics (QCD) and the masses of the electron, of the up quark and the down quark \cite{Damour10}. To translate the variation of a fundamental constant $C$ as a function of oscillations of the scalar field, we write it as $C(\phi)=C_0(1+d_C \varphi)$ with $C_0$ the value of the non-perturbed constant and $d_C$ the intensity of the coupling with the scalar field $\phi$.\\
\indent The current state of knowledge in astrophysics allows us to consider that dark matter is stable, self-gravitating 
and due to the velocity distribution in the galactic halo, this scalar field is a sum of stochastic scalar fields, giving a characteristic signature for its possible detection \cite{SavalleThese20}. For simplicity, we here approximate this field in the classical form $\phi=\phi_0 \text{cos}(\omega_m t)$ where $\omega_m=m_{\phi}c^2/\hbar$ is the associated angular frequency at the Compton wavelength \cite{Hees18} and $\phi_0$ is the amplitude related to the local dark matter density. In practice, the scalar field theoretically presents a spatial oscillation, but it is neglected due to the small size of the experiment \cite{Hees16, Damour10}. Thus, we expect to observe a variation of the aforementioned physical constants at this frequency, inducing a variation of the Bohr radius $a_0=\hbar/(\alpha c m_e)$. This effect will produce a change in the frequency of clocks and variations in the length of objects. It is not possible to measure the variation in length of an object with a material tool, like a ruler, since it would be affected by this variation. On the other hand, the speed of light is not modified by the presence of dark matter. From this, and because dark matter is not sensitive to electromagnetism, photonics is the perfect tool to seek to detect these variations, for instance through ultra-stable cavities.\\
\indent The cavity, made of two mirrors optically bonded to a rigid spacer, and the scalar field can be represented by a damped harmonic oscillator and a driving force \cite{Arvanitaki15,canuel_exploring_2018}, with $D$ the displacement of the mirrors, $Q$ the mechanical quality factor of the resonator and $\omega_n$ the frequency of the oscillator. The right term of the equation (\ref{oscharm}) is the driving force due to the scalar field, with $L_0$ the length of the resonator unaffected by the scalar field and $\epsilon_L=d_e+d_{m_e}$ the coupling parameter between the scalar field and the fine structure constant and the electron mass, that the experiment aims to determine.

\begin{equation} \label{oscharm}
\ddot{D}(t)+\frac{\omega_n}{Q}\dot{D}(t)+\omega_n^2 D(t)=-\epsilon_L L_0\omega_m^2\text{cos}(\omega_m t).
\end{equation}

By resolving this equation, the length of the cavity is written \cite{SavalleThese20}~:

\begin{equation} \label{varlong}
L(t)=L_0[1-\epsilon_L((1+\alpha)\text{cos}(\omega_mt)+\beta\text{sin}(\omega_mt))],
\end{equation}

and the coefficients $\alpha$ and $\beta$ are~:
\begin{equation}\label{coeffs}
\begin{split} 
\alpha=\sum^\infty_{i=1}\frac{8}{n^2\pi^2}\frac{Q_n^2\omega_m^2(\omega_n^2-\omega_m^2)}{Q_n^2(\omega_n^2-\omega_m^2)^2+\omega_m^2\omega_n^2} \\
\beta=\sum^\infty_{i=1}\frac{8}{n^2\pi^2}\frac{Q_n\omega_n\omega_m^3}{Q_n^2(\omega_n^2-\omega_m^2)^2+\omega_m^2\omega_n^2}, 
\end{split}
\end{equation}

with $Q_n=Q/n$ the quality factor of the mechanical resonator at the $n$-th harmonic \cite{Savalle21}.\\
From the length fluctuations in the equation (\ref{varlong}), the laser frequency oscillations can be estimated to~:
\begin{equation} \label{varfreq}
\frac{\delta\omega(t)}{\omega_0}=\epsilon_L[\varepsilon_c(1+\alpha)\text{cos}(\omega_mt)+\varepsilon_s\beta\text{sin}(\omega_mt)].
\end{equation}
The coefficients $\varepsilon_c$ and $\varepsilon_s$ depend on the optical properties of the cavity and are very close to 1 for our typical finesse and the scalar field frequency of interest ($f_m > 10$ kHz) \cite{Savalle21}. The mechanical resonance frequencies are given by $f_n=c_{s}/2L$ depending on the size and composition of the spacer, with $c_s$ the speed of sound in the spacer. The fundamental mode is generally a few tens of kHz.\\

\subsection{Effect of the scalar field on our silicon cavity}

\begin{figure}[h!]
	\begin{adjustwidth}{0cm}{0cm}
		\center
		\includegraphics[width=8cm]{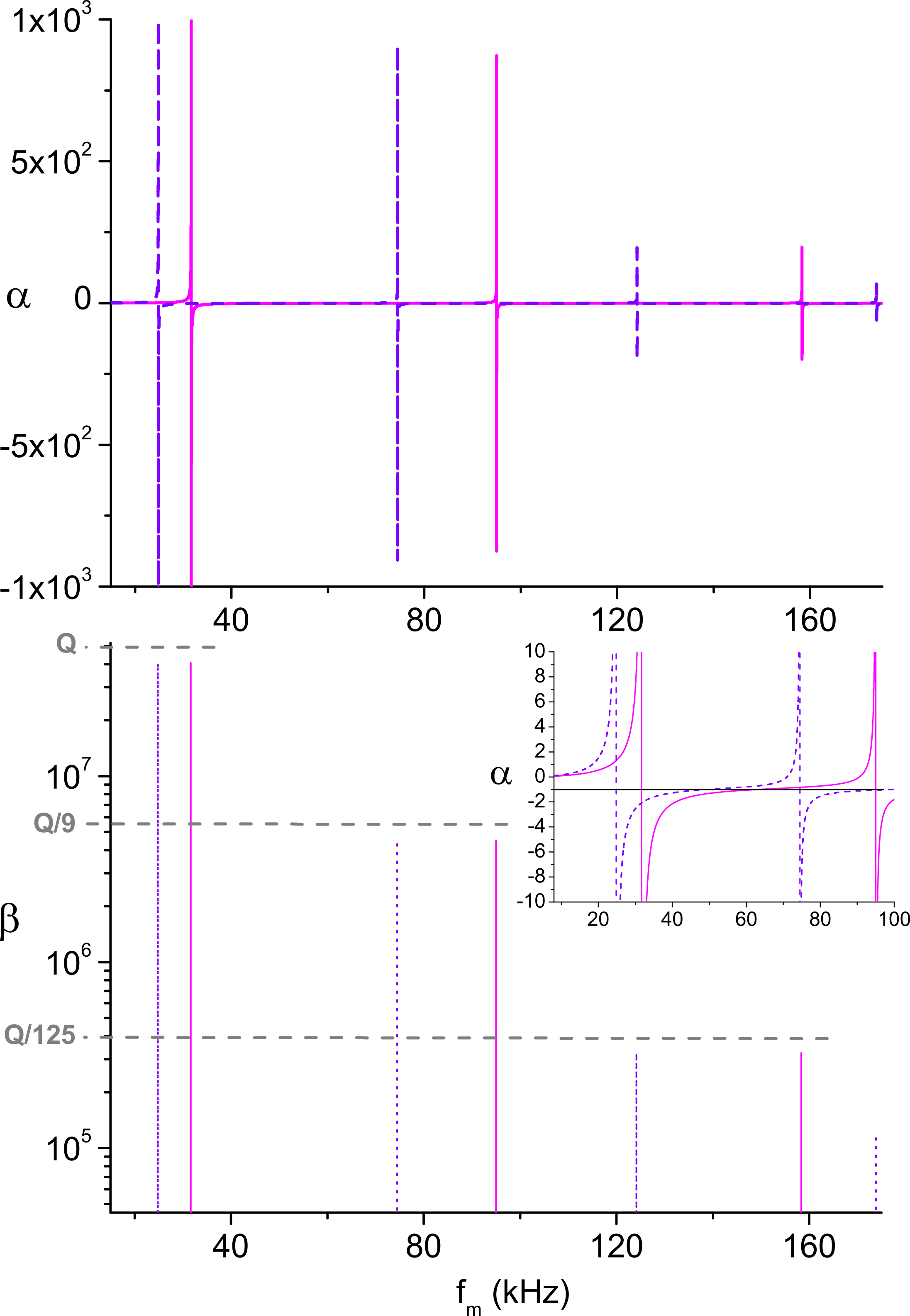}
	\end{adjustwidth}
	\caption{Plot of the $\alpha$ and $\beta$ coefficients for our 14 cm in purple continuous line. The dashed lines are the plot of the two coefficients for an 18 cm-long silicon cavity. In the insert, zoom on the $\alpha$ coefficient and the first harmonic. The horizontal black line shows the value $\alpha=-1$.}
	\label{resonances}
\end{figure}

\indent 
To estimate the speed of sound in the monocrystalline silicon spacer, one can use the relation~:

\begin{equation}
c_{s}=\sqrt{\frac{E}{\rho}}
\end{equation}

where $E$ is Young's modulus and $\rho$ is volume density. The volume density of crystalline silicon is $\rho=2330$ kg//m$^3$, but Young's modulus depends on the axis of the crystal. The spacers of our cavities are cut along the $[111]$ axis which has the largest Young's modulus, i.e. $E=185$ GPa. The longitudinal sound speed in the spacer is of $c_{s}=8910$ m/s, or about 70 \% greater than in the ULE glass.
For our 14 cm long cavity, we will have a first resonance frequency $\nu_{1} \approx 32$ kHz. The coefficient $\alpha$ from the equation (\ref{coeffs}) is plotted as a function of the scalar field frequency, for the 4 first odd harmonics and for the 14 cm cavity in silicon in purple continuous line (Figure~\ref{resonances}). The coefficient $\beta$ being very sharp around the resonance, it is essentially the term in (1+$\alpha$) which dominates out of resonances, in the equations (\ref{varlong}) and (\ref{varfreq}). When $f_m \ll f_n$, $\alpha$ and $\beta$ tend to 0, the cavity does not resonate but oscillates at the frequency of the scalar field. Conversely, when $f_m \gg f_n$, $\alpha$ tends towards $-1$ as can be seen in the insert of the Figure~\ref{resonances} in which the first harmonic of $\alpha$ as a function of $f_m $ is enlarged. We then have $l(t)=l_0$ and the cavity adopts a constant length if $f_m$ is not close to a resonance. Finally, at resonance, the cavity gains a factor $Q_n$ on its sensitivity.
On the lower panel of the Figure~\ref{resonances}, the coefficient $\beta$ decreases in $Q/n^3$ with the order of the harmonics, thus limiting the sensitivity of the instrument for high orders.

\begin{figure}[h!]
	\begin{adjustwidth}{0cm}{0cm}
		\center
		\includegraphics[width=8cm]{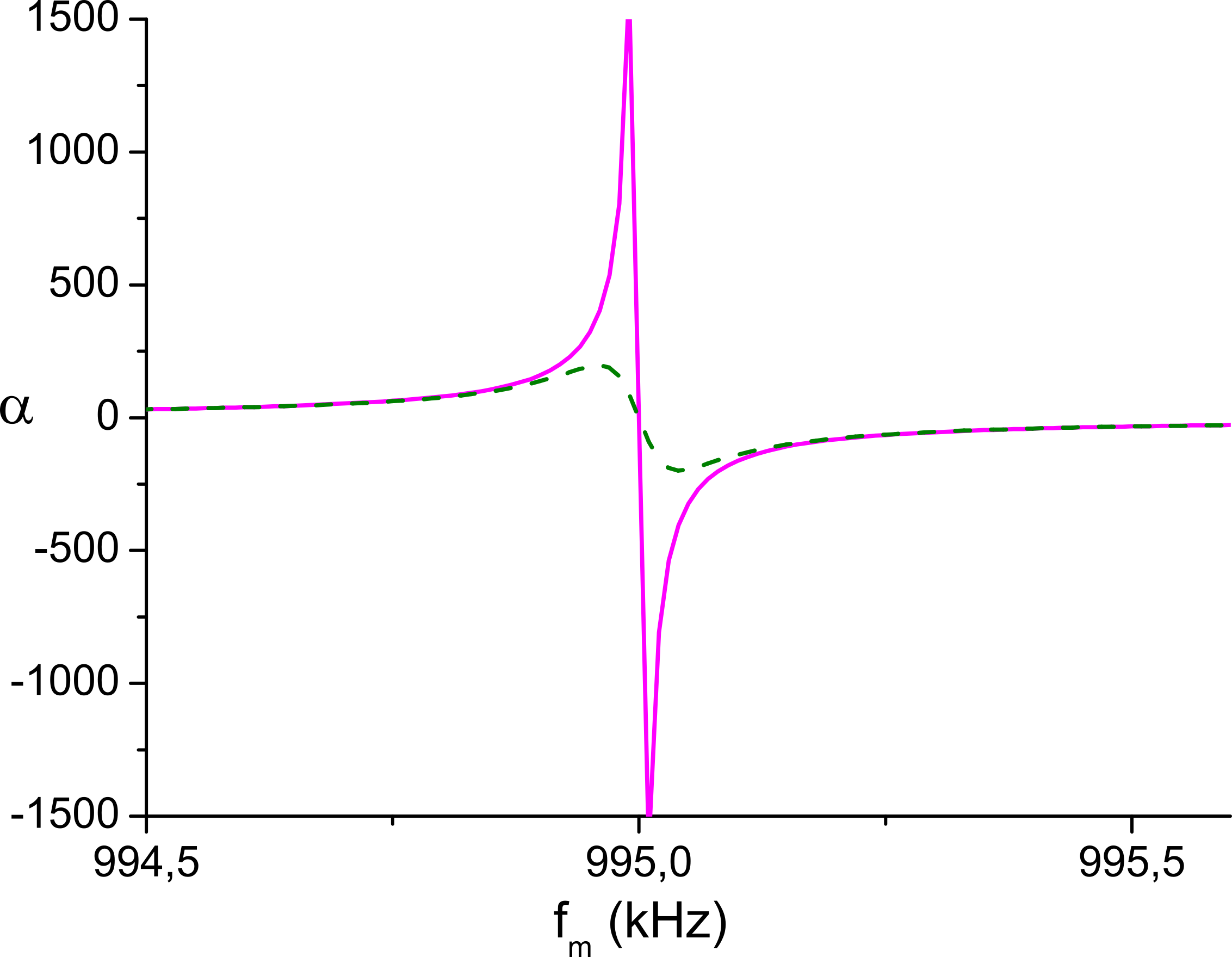}
	\end{adjustwidth}
	\caption{Plot of the $\alpha$ coefficient for a spacer in ULE (green dashed line) and a spacer in silicon (continuous purple line) for a resonance at the same frequency.}
	\label{Si-ULE}
\end{figure}

In addition, the quality factor of silicon allows for a significant increase the sensitivity. With the experiment carried out at SYRTE \cite{Savalle21}, the quality factor of ULE glass is $Q_{ULE}=6.1 \times 10^ {4}$ \cite{Millo09}, while with mono-crystalline silicon, the quality factor reaches $Q_{Si}=5 \times 10^{7}$, or almost 1000 times greater. As shown in the insert of the Figure~\ref{Si-ULE}, we therefore benefit from a much more sensitive instrument by the simple fact of using mono-crystalline silicon.

Since we also dispose of a 18 cm-long silicon cavity, we could use it for dark matter detection in the future. As seen on Figure (\ref{resonances}), using cavities of different lengths increases the peaks of sensitivity.\\
\indent The number of resonances that have to be considered depends on the bandwidth of the Pound-Drever-Hall servo loop. The piezo of the laser and the rapid correction of the AOM allows a bandwidth limited to about 200 kHz. However, we can significantly increase this bandwidth by using an EOM \cite{Kohlhaas12}, which will bring it to a value between 2 MHz and 4 MHz, thus covering a much wider scalar field energy range up to $\approx 10^{-8}$ eV.\\
\indent However, the ultra-stable laser alone cannot measure directly the frequency of the laser. It is necessary to do a frequency beat note between the frequency of the laser and another frequency reference to downconvert in the RF domain the frequency difference. The sensitivity of the test is different if the reference is a cavity, a maser, an atomic clock, etc \cite{Kennedy20}. Without another frequency reference, it is also possible to perform the detection test by interferometry \cite{Savalle21}. Regardless the architecture of the ultra-light dark matter experiment we are choosing, the silicon cavity is a powerful tool to detect scalar fields or impose new constraints on the coupling parameters $d_e$ and $d_{me}$.

\section{Conclusion}

We presented the status of the development of a 17 K cryogenic silicon ultra-stable cavity at FEMTO-ST. This is a 14-cm long single-crystal silicon cavity, made of a horizontal spacer and two optically contacted mirrors with single-crystal silicon substrates and amorphous dielectric coatings. The estimated thermal noise of such a cavity is expected to be at 3$\ \times\ $10$^{-17}$ and we have presented the different limitations that prevent us from reaching the state-of-the-art. While the fractional frequency instability due to the RAM is now lower than the thermal noise, we estimate that the vibrations and the optical power fluctuations are the two main remaining challenges. We already implemented a first optical power lock, stabilizing the optical power to the order of a few nanowatts. This servo loop is suitable for the RAM cancellation servo loop and is sufficient to decrease the optical power fractional frequency instability at the level of the thermal noise. 
Concerning the vibrations, the implementation of an active servo loop compensating them is also under investigation.\\
\indent Once these limitations are overcome, we plan to set up an ultra-light dark scalar field detection test, in an energy range close to $10^{-10}$ eV with this ultra-stable silicon cavity. Ultra-light scalar fields are possible dark matter candidates that couple to several fundamental constants, among them the electron mass and the fine structure constant, and thus affect the length of the cavity through oscillations of the Bohr radius. With the remarkable aimed frequency stability of the laser, we hope to bring new constraints on the knowledge of the coupling parameters between fundamental constants and dark scalar fields.

\section*{Acknowledgements}
This work was supported by the LABEX Cluster of Excellence FIRST-TF (ANR-10-LABX-48-01), within the Program "Investissements d'Avenir” operated by the French National Research Agency (ANR),  by the EquipeX Oscillator IMP (Grant ANR 11-EQPX-0033) and EIPHI Graduate school (Grant ANR-17-EURE-0002) projects.

\section*{References}

\bibliographystyle{iopart-num}

\bibliography{biblio}

\end{document}